\documentclass [a4paper,12pt]{article}
\usepackage {graphicx,amsmath}
\begin{document}
%\begin{flushright}
%November, 2010 
%\end{flushright}
%\vspace{-15mm}
\begin{center}
 {\Large\bf Schr\"{o}dinger equation and classical physics} \\[2mm]
           Milo\v{s} V. Lokaj\'{\i}\v{c}ek    
\footnote{e-mail: lokaj@fzu.cz} 
\\
{\it Institute of Physics of the AS CR, v.v.i., 18221 Prague 8, Czech Republic } \\
\end{center}

{\bf Abstract } \\
Any time-dependent solution of Schr\"{o}dinger equation may be always correlated to a solution of Hamilton equations or to a statistical combination of their solutions; only the set of corresponding solutions is somewhat smaller (due to existence of quantization). There is not any reason to the physical interpretation according to Copenhagen alternative as Bell's inequalities are valid in the classical physics only (and not in any alternative based on Schr\"{o}dinger equation). The advantage of Schr\"{o}dinger equation consists then in that it enables to represent directly the time evolution of a statistical distribution of classical initial states (which is usual in collision experiments). The Schr\"{o}dinger equation (without assumptions added by Bohr) may then represent the common physical theory for microscopic as well as macroscopic physical systems. However, together with the last possibility the solutions of Schr\"{o}dinger equation may be helpful also in analyzing the influence of other statistically distributed properties (e.g., spin orientations or space structures) of individual matter objects forming a corresponding physical system, which goes in principle beyond the classical physics. 
In any case, the contemporary quantum theory represents the phenomenological approximative description of some matter characteristics only, without providing any insight into quantum mechanism emergence. In such a case it is necessary to take into account more detailed properties at least of some involved objects.  \\ [10mm] 
%%%%%%%%%%%%%%%%%%%%%%

Two basic physical theories were being applied to physical reality in the end of 20th century: classical physics based on Hamilton equations and Copenhagen quantum mechanics based on the Schr\"{o}dinger equation. They represented quite different physical systems even if Schr\"{o}dinger \cite{schr} showed (in 1925) that the given equation 
 \begin{equation}
  i\hbar\frac{\partial}{\partial t}\psi(x,t)=H\psi(x,t), \;\;\;\;
     H=-\frac{\hbar^2}{2m}\triangle + V(x)   \label{schr}
\end{equation}
might describe the physical characteristics of particles exhibiting inertial motion if their positions and momenta were expressed with the help of corresponding time dependent function $\psi(x,t)$ as expectation values of corresponding operators 
\begin{equation}
    A(t) = \int\psi^*(x,t)\,A_{op}\,\psi(x,t)\,dx; \;\;    
          q^j_{op}=x_j,\;p^j_{op}=i\hbar\frac{\partial}{\partial x_j} .   \label{amp}
\end{equation} 
%The Schr\"{o}dinger equation has been denoted as wave equation and interpreted as describing a physical field, even if also trajectory interpretation was derived from Eq. (\ref{amp}).  

In the following we shall try to show the actual import of the Schr\"{o}dinger equation but also to introduce its contemporary fundamental limitation: its approximative phenomenological character. In the final part we shall attempt then to indicate how to extend our possibility to learn to know more about the quantum physical mechanism.   

In the first part we shall repeat shortly the story of quantum theory, stressing the individual critical points that have not been taken sufficiently into account in the past. First of all, it is necessary to call the attention to the basic fact that the Schr\"{o}dinger equation may be derived for the set of statistical combinations of Hamilton equation solutions when the given set has been limited by a suitable condition (e.g., by Boltzmann statistics); see \cite{hoyer,ioan}. It means that any function $\psi(x,t)$ may represent always a classical state or a statistical combination of such states; see also \cite{adv}, or \cite{mlok} where the problem of contemporary quantum theory has been summarized. 

Main difference between Schr\"{o}dinger equation and Hamilton equations system consists then in the fact that the set of physical states described by Schr\"{o}dinger equation is smaller, which follows from the existence of quantized states in closed systems only (as already mentioned). It is then necessary to distinguish the basic states (determined always by one Hamiltonian eigenfunction only) from other solutions of corresponding Schr\"{o}dinger equation. The basic states correspond to individual solutions of Hamilton equations (and represent pure states) while their superpositions correspond to their statistical combinations.

It is also important that the Schr\"{o}dinger equation is linear differential equation in time variable {\it{t}}. It means that the solutions corresponding to individual {\it{t}} values may be represented by vectors in a Hilbert space. However, when this Hilbert space is to retain the same physical interpretation (as in the classical physics) it must be extended in the way corresponding to given physical system. In principle, if the opposite time directions (e.g., in a two-particle system) are to be distinguished the given Hilbert space is to consist at least from two orthogonal Hilbert subspaces representing the incoming or outgoing states of these particles. The corresponding structure was described to a greater detail in two books of Lax and Phillips \cite{lax1,lax2}; see also \cite{mvl98}. 

Now it is possible to pass to the mentioned story.
The regular physical interpretation described in the preceding paragraph was deformed strongly when Bohr introduced his additional assumptions \cite{bohr1}: (i) the whole Hilbert space was required to consist of one subspace spanned on one set of basic states; (ii) all vectors of such a space were assumed to represent pure states. It was shown already by Pauli \cite{pauli} that in such a case one important contradiction existed in the given mathematical model. The corresponding Hamiltonian had to possess continuous spectrum in the interval \mbox{$E \in (-\infty,+\infty)$}. However, his argument remained at that time practically without any greater feedback. Later, it was shown by Susskind and Glogover \cite{suss} that the given Hilbert space was to be denoted as incomplete, too, as the exponential phase operator $e^{-i\Phi}$ was not unitary. Many attempts to solve these two shortages were then done in the second half of the 20th century. It is possible to say that the definite solution has been published in the beginning of this century \cite{kulo1,kulo2} when it has been shown that each of both the cases must be solved independently. In the first case the Hilbert space is to be extended according to the already mentioned proposal of Lax and Philips, while in the other case the given Hilbert space is to be extended further according to the proposal of Fajn \cite{fajn}. The total Hilbert space consists then of several mutually orthogonal subspaces, each of them being spanned on one set of Hamiltonian eigenfunctions:  
\begin{equation}
                H\psi_E(x) = E\psi_E(x);      \label{bas}
\end{equation} 
corresponding time-evolution trajectory passing through several subspaces.

Much more important criticism against the Copenhagen quantum mechanics of Bohr was brought, however, by Einstein \cite{ein} in 1935 who argued on the basis of a coincidence Gedankenexperiment that the given theory requires for two very distant matter objects to influence immediately one another (the phenomenon denoted later as entanglement). Einstein's critique (based in principle on the ontological approach to matter world) was refused by Bohr \cite{bohr} and also by the whole scientific community. It was accepted that the given interaction might exist between microscopic objects and two different physical theories were considered for microscopic or macroscopic physical worlds.
      
The certain change occurred in 1952 when Bohm \cite{bohm} showed that the original statement of von Neumann \cite{vonne} refusing the existence of hidden (i.e., local) variables in Bohr's quantum mechanics was not probably fulfilled. Two different physical interpretations were then considered in the microscopic world: the orthodox interpretation of Bohr involving the known quantum paradoxes and/or trajectory interpretation of the so called hidden-variable theory. 
Even if nobody analyzed the assumption bases of these two alternatives it was believed that it might be decided between them when Bell \cite{bell} derived in 1964 his famous inequalities       
\begin{equation}
             B = a_1b_1+a_2b_1+a_1b_2-a_2b_2 \leq 2 \, .    \label{ineq}
\end{equation}
It was assumed that they should have held in the framework of hidden-variable theory (and not in the Copenhagen alternative) for two pairs of simple probabilities in the slightly modified coincidence experiment 
                  \[   \left\|<--|^{\beta}---o---|^{\alpha}-->\right\|  \]
proposed originally by Einstein. The experiment of this type was then performed in 1982 (see \cite{asp}); two photons with opposite spins have been emitted in opposite directions by excited atom and the transmission probabilities through differently oriented polarizers in individual coincidence cases were being established.
   It was concluded that the inequalities of Bell were violated and the Copenhagen quantum mechanics was taken as the only theory of the microscopic physical world. 

However, to derive the given limit in inequalities (\ref{ineq})$\;$ Bell had to introduce the assumption requiring for the combination of four given probabilities (two pairs from opposite polarizers) to remain the same if one pair was interchanged and the other remained unchanged. It means that any dependence on polarizer deviations has been excluded.
Bell's inequalities might hold, therefore, in the classical physics only and not in any quantum alternative; the given problem having been explained to a greater detail in \cite{ml12}.
 The given inequalities were derived, of course, in other ways, too; see, e.g., Ref. \cite{clau}, where instead of one assumption of Bell several weaker assumptions have been made use of. When some assumptions have corresponded to a quantum alternative at least one of these assumptions has held always in the classical physics only (see, e.g., \cite{lk98}).

It means that the purely classical theory may be excluded on the basis of corresponding experimental data while any argument does not exist against the proper Schr\"{o}dinger equation. If one takes into account the existence of logical contradictions (interpreted as quantum paradoxes) in Copenhagen alternative the Schr\"{o}dinger equation (without assumptions added by Bohr) may be denoted as the theory being valid for the whole physical matter world, as the differences between quantum values at high energy values are practically unmeasurable; see also \cite{mlok} and \cite{mvl98}.  

It is, of course, possible to ask whether the lower set of physical states (i.e., the existence of quantization) is the only difference between the classical physics and Schr\"{o}dinger equation for which any of its solutions may be correlated to a state described by Hamilton equation (or to a statistical combination of such states). Certain advantage of Schr\"{o}dinger equation may be seen in that individual time dependent functions $\psi(x,t)$ may represent the measured results also in the cases when initial states are to be represented by some statistical combinations of different basic  states. However, in such a case
 also the influence of some other internal (not yet specified) states of individual matter objects may be involved that might be only hardly included into the system of Hamilton equations (e.g., the different spin orientations or internal space structures). 
However, in any case the narrow correlation between the Schroedinger equation and classical physics indicates that it is necessary to return again to the ontological approach to human knowledge on which all earlier successes of physical knowledge were based. 

On the other side it is necessary to introduce that the Schr\"{o}dinger equation represents the phenomenological description of some external characteristics of corresponding physical systems only and is not able (at least at the contemporary stage) to describe any mechanism leading to emergence of quantum states. One may demonstrate it on the hydrogen atom consisting of one proton and one electron. Such a stable object is to be formed when a slowly moving electron appears in the neighborhood of an oppositely charged proton. The electron is attracted to the proton but it is evident that  the quantum state can emerge only when a repulsive force exists between these two objects at the distance corresponding approximatively to proton dimension. This force is to be, of course, very short-ranged or contact. It means that the physical quantum states cannot be characterized by the values of Coulomb potential only. They should correspond in principle to the zero value of the sum of both the involved potentials (when also the mentioned contact force is characterized at least approximately by such a quantity).

It is, therefore, also the structure of free proton that should be responsible for final dimensions (and other characteristics) of the basic state of hydrogen atom. The study of corresponding physical characteristics should represent inseparable part of any quantum physics. The corresponding results may be obtained by studying the collisions of electrons with protons or the mutual collisions of protons at different energy values. However, one cannot be satisfied with the data obtained on the basis of mere phenomenological description only as it is usually done.  The models enabling to study the dependence on collision impact parameter must be made use of; see, e.g., the eikonal model applied to elastic p-p collisions at ISR energies \cite{01}. 

The given model has been  further generalized in \cite{02} where some more detailed information concerning the proton structure have been derived. According to these preliminary results the proton structure may be expected to be changeable and two (most frequent) proton structures of greatest dimensions may be responsible for the main part of differential $\;p\!-\!p\;$ cross section at lower scattering angles. They might be also responsible for the structure and dimensions of hydrogen atom. 
However, the maximum dimensions of these states are approximately $2f\!m$, which differs fundamentally from Bohr's value of hydrogen atom derived on the basis of Coulomb potential only. 

Before finishing it is necessary to stress that it is not more possible to represent the physical reality by some mathematical structures only. It is necessary to look for concrete ontological characteristics of corresponding matter objects even at the microscopic level. Only in such a way we may comprehend more about the internal structures at least of strongly interacting particles, which should be the main goal of contemporary quantum physics.

 In conclusion allow me to thank very much to my close collaborators Dr.~V.~Kundr\'{a}t and J.~Proch\'{a}zka for numerous discussions and critical comments.

{\footnotesize


\begin{thebibliography}{99}
\bibitem{schr}
E.Schr\"{o}dinger: Quantisierung als Eigenwertproblem; Ann. Phys.
{\bf 79}, 361-76; 489-527; {\bf 80}, 437-90; {\bf 81}, 109-39 (1926).
\bibitem{hoyer}
U.Hoyer:  Synthetische Quantentheorie;  Georg Olms Verlag, Hildesheim (2002). 
\bibitem{ioan}
H.Ioannidou: A new derivation of Schr\"{o}dinger equation;  Lett. al Nuovo Cim. {\bf 34},  453-8 (1982).  
\bibitem{adv}
M.V.Lokaj\'{\i}\v{c}ek: Schr\"{o}dinger equation, classical physics and Copenhagen quantum mechanics; New Advances in Physics {\bf 1}, No. 1, 69-77 (2007); see also /arxiv/quant-ph/0611176.
\bibitem{mlok} 
M.V.Lokaj\'{\i}\v{c}ek: Einstein-Bohr controversy after 75 years, its actual solution and consequences; in "Some applications of Quantum mechanics" (ed. M.R. Pahlavani), InTech Publisher, February 2012, pp. 409-424.
\bibitem{lax1}
P.D.Lax, R.S.Phillips: Scattering theory; Academic Press (1967).
\bibitem{lax2}
P.D.Lax, R.S.Phillips: Scattering theory for automorphic functions; Princeton (1976).
\bibitem{mvl98}
M.V.Lokaj\'{\i}\v{c}ek:Realistic theory of microscopic phenomena; a new solution of hidden-variable proble; /arXiv:quant-ph/9811030 (1998)
\bibitem{bohr1}
N.Bohr: The quantum postulate and the development of atomic
theory; Nature  121, 580-90 (1928).
\bibitem{pauli}
W.Pauli: Die allgemeinen Prinzipien der Wellenmechanik; Handbuch der Physik 
{\bf XXIV}, Springer, Berlin, p. 140 (1933).
\bibitem{suss}
L.Susskind, J.Glogover: Quantum mechanical phase and time operator;  Physics (Long Island City, N.Y.)  {\bf 1},  49-61 (1964).
\bibitem{kulo1}
P.Kundr\'{a}t, M.Lokaj\'{\i}\v{c}ek: Three-dimensional harmonic oscillator and time evolution in quantum mechanics; Phys. Rev. A {\bf 67}, art. 012104 (2003).
\bibitem{kulo2}
P.Kundr\'{a}t, M.Lokaj\'{\i}\v{c}ek: Irreversible time flow and Hilbert space structure; New Research in Quantum Physics (eds. Vl.Krasnoholovets, F.Columbus), Nova Science Publishers, Inc.,  pp. 17-41 (2004).
\bibitem{fajn}
V.Fajn: Quantum harmonic oscillator in phase representation and
the uncertainty relation between the number of quanta and the
phase (in Russian);  J. Exp. Theor. Phys.  {\bf 52}, 1544-8 (1967).
\bibitem{ein}
A.Einstein, B.Podolsky, N.Rosen: Can quantum-mechanical description of physical reality be considered complete?; Phys. Rev.  47,  777-80 (1935).
\bibitem{bohr}
N.Bohr: Can quantum-mechanical description of physical reality be
considered complete?;   Phys. Rev.  48, 696-702 (1935).
 \bibitem{bohm}
D.Bohm: A suggested interpretation of the quantum theory in terms
of "hidden variables"; Phys. Rev. {\bf 85}, 180-93 (1952).         
\bibitem{vonne}
J.von Neumann:  Mathematische Grundlagen der Quantenmechanik; Springer (1932).
\bibitem{bell}
J.S.Bell: On the Einstein Podolsky Rosen paradox; Physics  1, 195-200 (1964).
\bibitem{asp}
A.Aspect, P.Grangier, G.Roger: Experimental realization of
Einstein-Podolsky-Rosen-Bohm Gedankenexperiment: A new violation
of Bell's inequalities; Phys. Rev. Lett.  49, 91-4 (1982).
\bibitem{ml12}
M.V.Lokaj\'{\i}\v{c}ek: The assumption in Bell's inequalities and entanglement problem; J. Comp. Theor. Nanosci. (accepted for publication); see also  /arXiv:1108.0922
\bibitem{clau}
J.F.Clauser, A.Shimony: Bell's theorem: experimental tests and
implications; Rep. Prog. Phys. {\bf 41}, 1881-91 (1978).
\bibitem{lk98}
M.V.Lokaj\'{\i}\v{c}ek: Locality problem, Bell's inequalities and EPR experiments; /arXiv:quant-ph/9808005 (1998).
\bibitem{01}
V. Kundr\'{a}t, M. V. Lokaj\'{\i}\v{c}ek:  High-energy elastic scattering amplitude of unpolarized and charged hadrons; Z. Phys. C  63 , 619-29 (1994) 
\bibitem{02}
M. V. Lokaj\'{\i}\v{c}ek, V. Kundr\'{a}t: Elastic pp scattering and the internal structure of colliding protons;  /arXiv:0909.3199[hep-ph] (2009)

\end{thebibliography}
\end{document}